# Determination of the spin Hall angle by the inverse spin Hall effect, device level ferromagnetic resonance, and spin torque ferromagnetic resonance: a comparison of methods

Ranen Ben-Shalom[1], Nirel Bernstein[1], See-Hun Yang[2], Amir Capua[1]*

[1]Department of Applied Physics, The Hebrew University of Jerusalem, Jerusalem 91904, Israel

[2]IBM Research Division, Almaden Research Center, 650 Harry Rd., San Jose, California 95120, USA

*e-mail: amir.capua@mail.huji.ac.il

**Abstract:**

**The spin torque ferromagnetic resonance (STFMR) is one of the popular methods for measurement of the spin Hall angle, $\theta_{SH}$. However, in order to accurately determine $\theta_{SH}$ from STFMR measurements, the acquired data must be carefully analyzed: The resonance linewidth should be determined to an accuracy of a fraction of an Oe, while the dynamical interaction leading to the measured response consists of the conventional field-induced ferromagnetic resonance (FMR), spin-torque induced FMR, and of the inverse spin Hall effect (ISHE). Additionally, the signal often deteriorates when DC current is passed through the device. In this work we compare the STFMR method with two other FMR-based methods that are used to extract $\theta_{SH}$. The first is a device-level FMR and the second is based on the ISHE. We identify artefacts that are caused by the noise floor of the instrumentation that make the measurement of $\theta_{SH}$ illusive even when the signal to noise ratio seems to be reasonable. Additionally, we estimate a 10% error in $\theta_{SH}$ that results from neglecting the magnetic anisotropies as in conventional measurements. Overall, we find the STFMR to be the most robust of the three methods despite the complexity of the interaction taking place therein. The conclusions of our work lead to a more accurate determination of $\theta_{SH}$ and will assist in the search of novel materials for energy efficient spin-based applications.**

Efficient generation of spin currents is key to spintronics. Over the last decade the majority of the efforts was focused on finding new physics for polarizing the transported spins. Such an example is the intensive study of the spin Hall effect (SHE) [1,2] that relies on the spin orbit coupling (SOC) or on non-trivial topologies of the band structure [3]. In addition to the challenges in generating spin currents, their accurate quantification is not straight forward. For example, in Pt, which is one of the main materials used to investigate the SHE, a large variation in the SHE efficiency was reported [4-7].

The charge to spin current conversion rate is referred to as the spin Hall angle [8], $\theta_{SH}$, and is defined by $\theta_{SH} = 2eJ_S/\hbar J_C$ where $e$ is the electron charge, $\hbar$ is Planck's constant, $J_C$ is the charge current density, and $J_S$ is the spin current density. Measurement of $\theta_{SH}$ can be carried out by various methods such as the cavity ferromagnetic resonance (FMR) [9,10], spin pumping [11-15], optical FMR [16,17], anomalous and planar Hall effect [18-21], spin torque ferromagnetic resonance (STFMR) [5,22-25], and nonlocal spin transport [8,26-28]. Among these methods, the STFMR is most commonly used due to its simplicity and applicability. In the STFMR, an AC charge current is injected into a bilayer of a ferromagnet (FM)/heavy metal (HM) and excites spin precessions. Due to the anisotropic magneto resistance (AMR) [29], a rectified DC voltage builds up. Accordingly, the measured FMR response stems from three different processes: 1) SHE in the HM that generates AC spin currents that are responsible for an antisymmetric lineshape. 2) A symmetric lineshape that stems from the stray AC Oersted field generated by the current passing in the HM. 3) An additional symmetric resonance response that originates in spin pumping [30] by the precessing magnetization $\vec{M}$ [22] in the FM layer. Therefore, analysis of the measured lineshape is not trivial and requires separating the different contributions, each resolved to a fraction of an Oe. This, in turn, usually requires solving a five-parameter lineshape fitting optimization problem [24].

In this work, we compare the STFMR technique with the inverse SHE (ISHE) [31] and device-level ferromagnetic resonance (DLFMR) methods for measurement of $\theta_{SH}$. All techniques are applied on the same device. Each method has relative advantages such as the elimination of the spin transfer torque (STT) [22] or in the simplicity of the analysis of the measured responses. Our study shows that the anisotropies that are usually neglected, e.g. the shape anisotropy, may play a considerable role [32]. However, we show that the conventional optimization algorithm used to analyze the STFMR response is robust. In addition, our measurements indicate a significant electrical background noise that affects all methods. This noise often deteriorates the ISHE and DLFMR signals and in the worst case leads to unnoticeable measurement artefacts that result in erroneous values of $\theta_{SH}$. We conclude that despite the relative advantages of the ISHE and DLFMR methods and the challenge of analyzing the STFMR response the STFMR is a more robust and reliable technique.

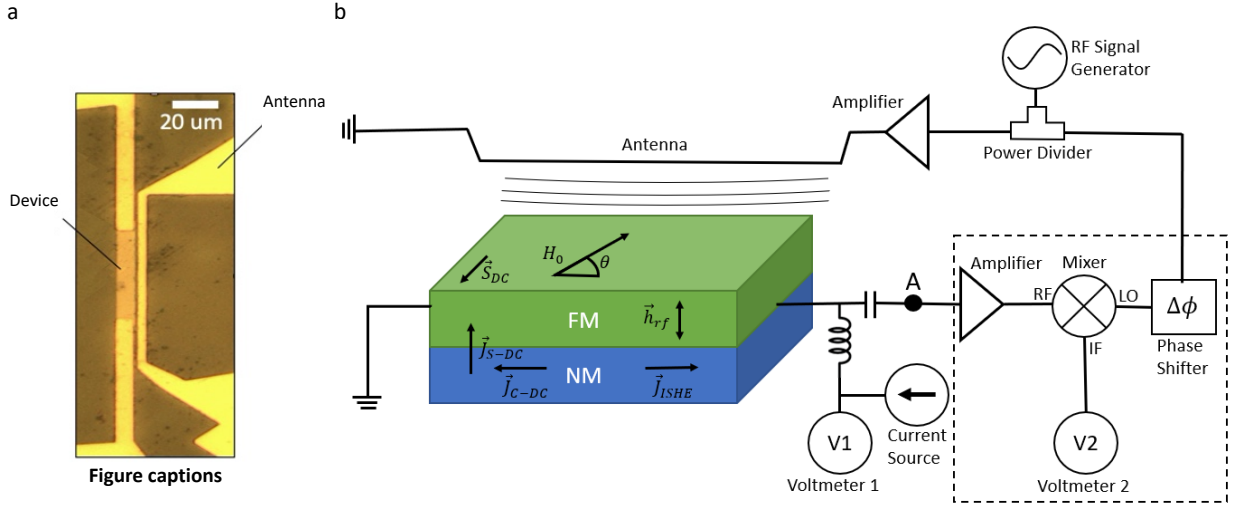

**Fig. 1. Experimental setup. (a) Patterned SHE device with a micro-antenna. (b) Schematic of the measurement setup that includes the STFMR, ISHE, and DLFMR.**

The experimental system is presented in Fig. 1 and can be converted between the STFMR, ISHE, and DLFMR modes of operation so that all methods are applied on the same device. In the STFMR a microwave signal is generated by the RF signal generator and is driven into the device by connecting the RF signal generator to node 'A' indicated in the figure. The injected AC current drives the FMR in the device in the presence of an externally applied DC magnetic field, $H_{ext}$. The oscillatory driving torque stems from the Oersted stray field and the AC STT that is generated by the SHE. Due to the AMR and the magnetization precessions, the AC current is rectified and a DC voltage builds up on the device. In addition, the precessions of $\vec{M}$ pump spin current by the spin pumping mechanism [30] into the HM which is converted into a DC charge current by the ISHE. Once the FMR is excited a DC charge current, $I_0$, that is converted into a DC spin current is injected and exerts an anti-damping torque that is reflected on the resonance linewidth from which $\theta_{SH}$ is extracted in the usual manner [22]. The DC voltage building upon the device in the STFMR configuration is given by $V_{STFMR} = I_0 R_0 + I_{ISHE}(H_{ext}) \cdot R_0 + \frac{1}{2} \cdot i_0 \cdot (\Delta R_{Oe}(H_{ext}) + \Delta R_{STT}(H_{ext}))$ where $i_0$, $R_0$, $I_{ISHE}$, $\Delta R_{Oe}$, and $\Delta R_{STT}$ are the amplitude of AC charge current, the DC resistance, the DC ISHE charge current that is pumped by $\vec{M}$, the amplitude of the AC resistance that originates in the AC Oersted excitation, and the amplitude of the AC resistance that originates from the STT, respectively. Both $\Delta R_{Oe}$ and $\Delta R_{STT}$ stem from the AMR. To extract the resonance linewidth, following a data fitting procedure, the measured resonances are decomposed to the symmetric and antisymmetric components that originate in the Oersted and STT excitations. The fitting process requires solving a five-parameter optimization problem of the following expression:

$$V_{STFMR} = A \frac{\Delta H^2}{(H_{ext}-H_{res})^2+\Delta H^2} + B \frac{\Delta H^2}{(H_{ext}-H_{res})^2+\Delta H^2} \frac{H_{ext}-H_{res}}{\Delta H} + C \qquad (1)$$

In Eq. 1 $H_{res}$ is the resonance field, $\Delta H$ is the resonance width, and A, B, and C are the coefficients of the symmetric, antisymmetric, and DC background level, respectively. The contribution of the ISHE in $V_{STFMR}$ is neglected in our analysis [22]. Our experimental results below indicate that this approximation is valid, although according to Ref. [33] the ISHE contribution was found to be significant.

In the other two configurations of the DLFMR and ISHE, the AC current is driven into a micro-antenna that is patterned in proximity to the device (Fig. 1) rather than being injected into it. In this manner we excite the FMR without the contribution of the AC STT. Therefore, the dynamical processes taking place in the ISHE and the DLFMR measurements are greatly simplified as compared to the STFMR. Consequently, the resonance responses of the ISHE and DLFMR are purely symmetric and in the data fitting procedure the optimization algorithm is applied to solve a four-parameter problem.

The ISHE and DLFMR differ in the manner at which the FMR signal is read. In the ISHE spin pumping induces a DC voltage signal by the ISHE in the HM layer that is probed. The measured voltage is given by $V_{ISHE} = I_0 R_0 + I_{ISHE}(H_{ext}) \cdot R_0$. In the DLFMR configuration the FMR signal is read at the fundamental RF harmonic. Due to the AMR effect, the resistance of the device is modulated at the RF frequency and in the presence of the applied DC current a voltage signal is generated at the fundamental frequency. A homodyne detection scheme using an RF mixer is then used to down-convert this signal and read it as illustrated in Fig. 1. The measured DLFMR signal is expressed by $V_{DLFMR} = (I_0 + I_{ISHE}(H_{ext})) \cdot \Delta R_{Oe}(H_{ext})$. The DLFMR signal contains also an ISHE contribution, $I_{ISHE}$, however, it is three orders of magnitude weaker than $I_0$. To further eliminate ISHE component, in the experiment we used a lock-in detection scheme in which $I_0$ was modulated.

In order to model the dynamics with a minimal set of approximations we derived the theoretical model based on the Smit-Beljiers-Suhl formalism [17,34,35]. Accordingly, the Landau-Lifshitz-Gilbert equation in spherical coordinates becomes:

$$\dot{\theta} = \gamma H_\phi, \quad sin\theta \dot{\phi} = -\gamma H_\theta \qquad (2)$$

Here, $\gamma$ is the gyromagnetic ratio while $H_\theta$ and $H_\phi$ are the effective fields along the polar and azimuthal unit vectors, respectively. These originate from the conserving energies, $E_{conserv.}$, that account for $H_{ext}$, the demagnetizing field, the crystalline anisotropy, the microwave field, and the field-like spin orbit torque (SOT): $E_{conserv.} = -\vec{M} \cdot (\vec{H}_{ext} + \vec{H}_{Oe} + \vec{h}_{Oe}) - 12\vec{M} \cdot \vec{H}_D - K_u \sin^2\theta - \vec{M} \cdot \vec{h}_{rf} - H_{FL}\vec{M} \cdot \hat{s}$. $\vec{H}_{Oe}, \vec{h}_{Oe}$, $\vec{H}_D, \vec{h}_{rf}, \vec{H}_{FL}, \hat{s}$, and $K_u$ are the stray DC Oersted field stemming from the HM, the Oersted field generated by AC current, the demagnetization field, the RF field of amplitude $h_{rf}$, the field-like STT, the spin polarization of the injected spins, and the magnetocrystalline anisotropy constant, respectively. In addition, $H_\theta$ and $H_\phi$ also stem from the nonconserving energies, $E_{nonconserv.}$ that account for the magnetic Gilbert damping losses and the damping-like SOT: $E_{nonconserv.} = (\alpha/2\gamma M_s)\dot{\vec{M}}^2 + H_{SHE}\dot{\vec{M}} \cdot (\vec{M} \times \hat{s})$ where $\alpha$ is the Gilbert damping and $M_s$ is the saturation magnetization. $H_\theta$ and $H_\phi$ are then determined from the effective field, $\vec{H}_{eff}$, that is given by $\vec{H}_{eff} = -\nabla_{\vec{M}} E_{conserv.} - \nabla_{\dot{\vec{M}}} E_{nonconserv.}$. The spin Hall parameter $H_{SHE}$ is given by $\hbar\theta_{SH}J_C/2eM_s t$ where $t$ is the thickness of the ferromagnetic layer. $E_{conserv.}$ and $E_{nonconserv.}$ have units of energy density and energy density flow.

Solving for small deviations, $(\Delta\theta, \Delta\phi)$ around the equilibrium state, $(\theta_0, \phi_0)$, in phasor-space we find the in-plane component of $\vec{M}$ at the angular frequency $\omega$ which is given by:

$$\Delta\phi = \frac{\gamma}{(1+\alpha^2)\sin\theta_0} \frac{(\omega^2-\omega_0^2)+i\omega\Delta\omega}{(\omega^2-\omega_0^2)^2+\omega^2\Delta\omega^2}\left[(h_\theta + \alpha h_\phi)i\omega + \frac{\gamma}{M_s}\left(\frac{1}{\sin\theta_0}h_\theta\zeta_{\phi_0\theta_0} + h_\phi\zeta_{\theta_0\theta_0}\right)\right] \quad (3)$$

with:

$$\omega_0^2 = \frac{\gamma^2}{M_s^2(1+\alpha^2)\sin^2\theta_0}\left[\zeta_{\phi_0\phi_0}\zeta_{\theta_0\theta_0} - \zeta_{\phi_0\theta_0}\zeta_{\theta_0\phi_0}\right]$$

$$\Delta\omega = \frac{\gamma}{M_s(1+\alpha^2)\sin\theta_0}\left[\alpha\sin\theta_0\zeta_{\theta_0\theta_0} + \frac{\alpha}{\sin\theta_0}\zeta_{\phi_0\phi_0} + \zeta_{\phi_0\theta_0} - \zeta_{\theta_0\phi_0}\right]$$

And $\zeta_{ij}$ defined by

$$\zeta_{\theta_0\phi_0} = E_{\theta_0\phi_0}, \quad \zeta_{\phi_0\theta_0} = E_{\phi_0\theta_0} + M_s H_{SHE}(s_x\cos\phi_0 - s_y\sin\phi_0)$$

$$\zeta_{\theta_0\theta_0} = E_{\theta_0\theta_0} + M_s H_{SHE}\sin\theta_0(-s_x\cos\theta_0\sin\phi_0 + s_y\cos\theta_0\cos\phi_0)$$

$$\zeta_{\phi_0\phi_0} = E_{\phi_0\phi_0} + M_s H_{SHE}\sin\theta_0(s_x\sin\theta_0\cos\phi_0 + s_y\sin\theta_0\sin\phi_0 - s_z\cos\theta_0)$$

In Eq. (3) $\omega_0$ is the resonance frequency, $\Delta\omega$ is the resonance linewidth, and $h_\theta$ and $h_\phi$ are the polar and azimuthal components of the RF excitation. $s_x, s_y$, and $s_z$ are the projection of $\hat{s}$ on the cartesian axis. $E_{\phi_0\phi_0}, E_{\phi_0\theta_0}, E_{\theta_0\phi_0}$, and $E_{\theta_0\theta_0}$ are the second order derivatives of $(E_{conserv.} + E_{nonconserv.})$, with respect to $\theta$ and $\phi$. The equilibrium state was evaluated without the contribution of the SOT.

To compare the methods we used the well-studied Pt based system: 60 Pt/5 Py/60 Co/20 TaN (units in Å). The film was grown by magnetron sputtering and patterned to a 5 X 40 $\mu m$ device. The Au micro-antenna of 100 $nm$ thickness and 2 $\mu m$ width was deposited at a distance of 1.5 $\mu m$. The AC current density driven into the device in the STFMR experiment was $10 \cdot 10^{12}\ A/m^2$ and $1.4 \cdot 10^{12}\ A/m^2$ into micro-antenna in the ISHE and DLFMR experiments. To maximize the signals, in the STFMR and DLFMR experiments the current was injected at 45° with respect to $H_{ext}$ and at 90° in the ISHE experiment.

The frequency dispersion relation and linewidth vs. frequency as obtained from the STFMR method are presented in Fig. 2. From these measurements the magnetic parameters of the sample were extracted using Kittel's formula, $f_0 = \gamma/2\pi \cdot \sqrt{H_{res}(H_{res} + 4\pi M_s)}$, with $f_0$ being the resonance frequency, and the frequency dependent linewidth, $\Delta H = 2\alpha/\gamma \cdot \omega + \Delta H_{IH}$, with $\Delta H_{IH}$ being the inhomogeneous broadening. Accordingly, the values of $\alpha = 0.0048$, $M_s = 1384\ emu/cm^3$, and $\Delta H_{IH} = 16\ Oe$ were extracted.

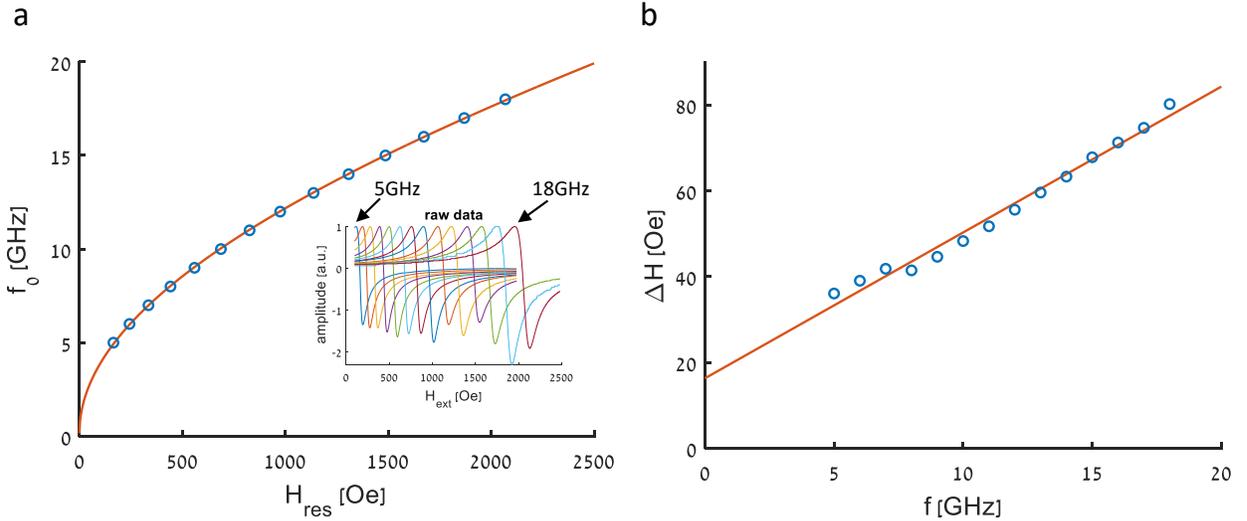

**Fig. 2. (a) STFMR Frequency-field dispersion relation. Open blue circles indicate the measurement. Solid red line indicates Kittel's formula. Inset illustrates the measured STFMR responses for 5 - 18 GHz. (b) Resonance linewidth as a function of frequency. Solid red line indicates the linear fit.**

The measurement of $\theta_{SH}$ using the STFMR and ISHE experiments are presented in Fig. 3. The experiments were carried out at 8 GHz. The figure presents $\Delta H$ and $H_{res}$ as function of $J_C$. Fig. 3(a) and 3(b) present the STFMR results. The shift of $H_{res}$ with $J_C$ originates from the stray DC Oersted field stemming from the DC current and agrees well with a calculation of the expected stray field. The linewidth measurements were fitted using Eq. (3) from which a $\theta_{SH}$ of 23.1% was extracted. This value is within the range of previously measured $\theta_{SH}$ values [4,17,36,37]. Interestingly, it is readily seen that the noise level increases as $|J_C|$ is increased. This behavior is marked by the shaded area in the figure and is typical of an intrinsic Johnson noise and the shot noise limited processes. These processes take place either in the device or the measurement instrumentation, or both. This noise limits the maximal DC current that can be injected. A similar behavior was also seen in Ref. [38]. There, it was shown that this limitation can be overcome by optically probing $\vec{M}$.

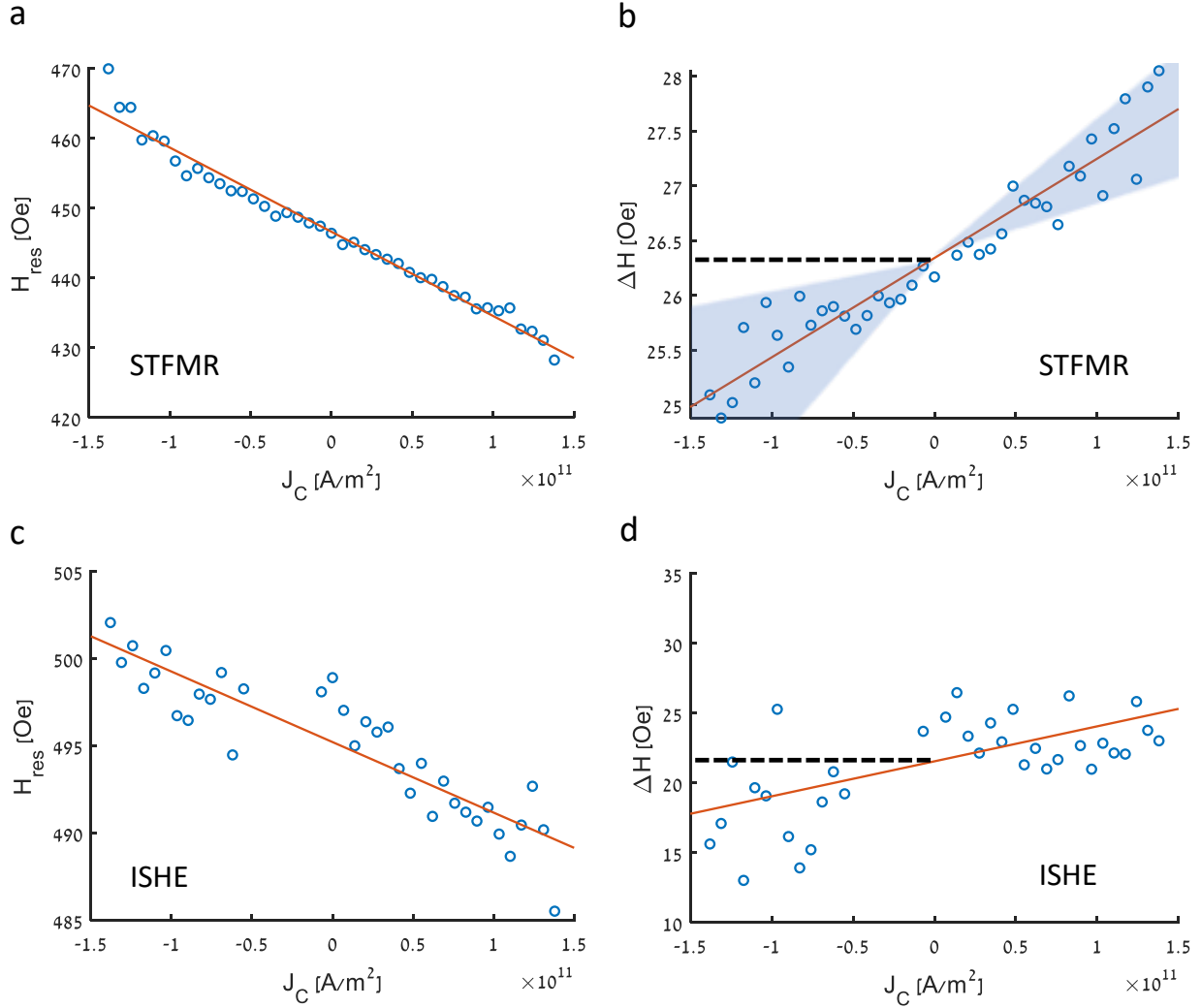

Fig. 3. (a) $H_{res}$ as a function of $J_C$ for the STFMR measurements. (b) $\Delta H$ as a function of $J_C$ for the STFMR measurements. Shaded blue area guides the eye to the noise level. Black dashed line indicates $\Delta H$ at $J_C = 0$. (c) $H_{res}$ as a function of $J_C$ for the ISHE measurements. (d) $\Delta H$ as a function of $J_C$ for the ISHE measurements. Black dashed line marks $\Delta H$ for $J_C = 0$. In (a) - (d) the open blue circles indicate the measurement and solid red lines indicate the linear fits. Data presented at 8 GHz.

The overall broadening that results from the anti-damping torque is only of a few Oe. Therefore, in order to extract $\theta_{SH}$, the linewidth should be determined to the accuracy of a fraction of an Oe. However, since the optimization algorithm is applied on a multi-parameter problem, five in our case, it is possible that it would converge into a local minimum. Therefore, we tested the convergence of the algorithm. This was carried out by providing the algorithm with initial conditions that are far off from the actual solution. We found that a shift larger than 100 Oe in the resonance frequency and 50 Oe in the linewidth was the limit after which the optimization algorithm converged to a wrong solution. This illustrates that the fitting algorithm is robust.

The dependence of $H_{res}$ and $\Delta H$ on the injected DC current in the ISHE measurement are shown in Fig. 3(c) and 3(d). The ISHE measurements are significantly noisier. Consequently, a wrong value of $\theta_{SH} = 46.8$ that is larger by a factor of two as compared to the STFMR method was measured. Despite the noise, we can readily see that the shift in $H_{res}$ as function of $J_C$ is much smaller than in the STFMR measurement. In the STFMR the overall shift is ~ 35 Oe while in the ISHE the shift is only ~ 11 Oe. Since the STFMR measurements are carried out at 45° and the ISHE measurements at 90° we can attribute the difference to the anisotropy fields which are usually neglected. However, our model which includes the anisotropy fields could not fully account for the differences and we believe there is an additional contribution such as the existence of magnetic domains. From our model, we found that neglecting the anisotropy fields results in an error of ~10% in $\theta_{SH}$.

The DLFMR configuration is the most direct measurement of the FMR. The FMR is excited externally by the micro-antenna and is read directly on the device using a homodyne detection scheme. In contrast to the other configurations, in the DLFMR the signal arises only from the AC resistance that is modulated by the AMR. Additionally, the homodyne detection scheme eliminates the DC noise which is responsible for a significant portion of the noise as shown above. For these reasons we expect the DLFMR experiment to be more accurate. In reality this is not the case and the DLFMR measurement (Fig. 4(a)) was noisy to the extent that we could not extract $\theta_{SH}$ reliably. We speculate that the noise originates from thermal effects in the RF mixer.

In order to compare the STFMR with the ISHE and DLFMR methods we plot only the symmetric part of the STFMR measurement that originates in the Oersted field. The symmetric part of the STFMR response has the same functional form as the DLFMR and ISHE lineshapes. A comparison of all lineshapes overlaid on top of each other is presented in Fig. 4. Each trace was normalized to its maximum. The ISHE resonance frequency is higher by ~ 50 Oe as compared to the STFMR, and DLFMR. This stems from the difference in the direction at which $I_0$ is applied with respect to $H_{ext}$. This behavior demonstrates once more that the anisotropies are not as negligible as conceived. The shift in the resonance field resulting from the different angle of applied $H_{ext}$ is also seen in the STFMR and ISHE measurements of $H_{res}$ as function of $J_C$ in Fig. 3.

Fig. 4(a) shows that the DLFMR and ISHE are much noisier as compared to the STFMR. At the origin of the different noise levels of the ISHE and the DLFMR lies the physical mechanism used to read out the signal. In the ISHE, the signal originates from the spin pumping and in the DLFMR and STFMR it originates in the AMR. These differences translate to a STFMR signal that is 26 times stronger than the ISHE signal and, therefore, lead to a higher signal to noise ratio (SNR). For this reason, conclude that the contribution of the ISHE is negligible in the STFMR experiment and can be neglected. In the DLFMR the signal originates in the AMR as in the STFMR. Hence, the magnitude of the signals readout is similar. In the DLFMR, the absolute noise level is higher than in the STFMR because of the mixer and possibly impedance mismatching effects that are not present in the DC measurements. Therefore, the SNR is significantly lower than in the STFMR measurement. To understand how the measurement is affected by the poor SNR we simulated a Lorentzian resonance function that is submerged in a background noise level as shown in Fig. 4(b). We then fitted the resultant signal to an ideal Lorentzian. It is seen that the fitted resonance is narrower than the real resonance. The narrower linewidth of the ISHE and DLFMR is also seen in our measurements when the ISHE lineshape is overlaid on the STFMR and DLFMR lineshapes as illustrated by the black dashed line in Fig. 4(a). Clearly, the DLFMR and ISHE resonances appear narrower than the STFMR resonance indicating the significant contribution of the noise. We evaluate the ISHE noise floor to be $0.8\mu V$ and the DLFMR noise floor to be $140\mu V$. Both are 7% of the resonance peak. We can see the same effect in the linewidth measurements of the STFMR and ISHE in Fig. 3. The absolute width of the ISHE spectrum is smaller than the STFMR width as indicated by the vertical black dashed lines of Fig. 3(b) and 3(d). This illusive artifact illustrates the challenges of reliably extracting $\theta_{SH}$.

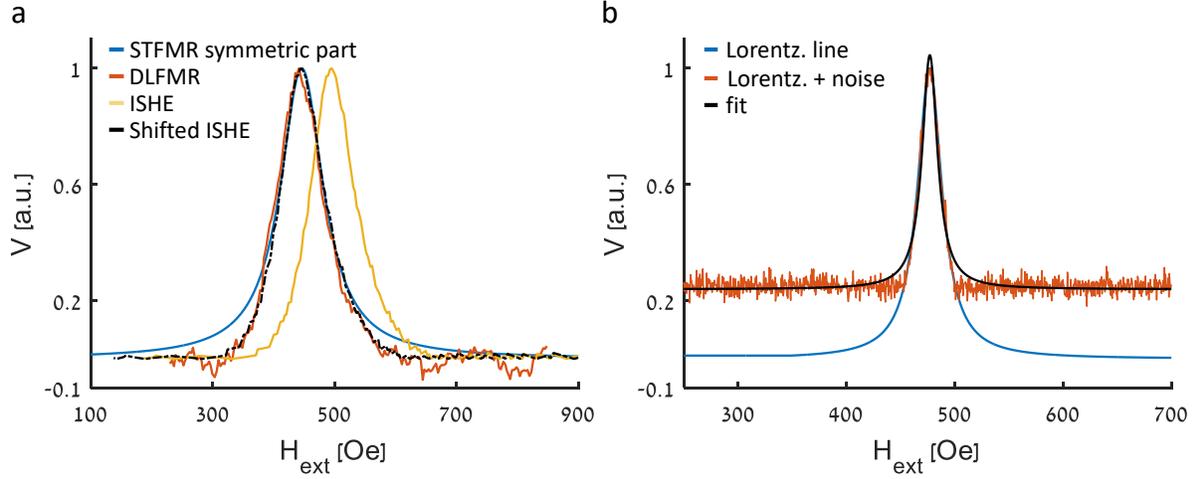

**Fig. 4. (a) Symmetric part of the STFMR measurement (blue) together with the responses of the DLFMR (red) and ISHE (yellow) measurements and the shifted ISHE response (dashed black). Traces are normalized to the peak value. (b) Noise simulation. Blue solid line indicates an ideal Lorentzian lineshape. Red solid line illustrates the calculated Lorentzian lineshape together with a simulated noise floor. Black solid line is a fit of the simulated noisy Lorentzian to a noise-free response.**

To summarize, in this work we have examined the intricate details of the commonly used STFMR technique and compared it to the ISHE and DLFMR methods that sense more directly the FMR response in the presence of spin currents. We found that despite the physical complexity of the dynamical interaction in the STFMR, it is more robust than the ISHE and DLFMR. We showed that it is necessary to consider the anisotropies, e.g. the shape anisotropy in the device to resolve more accurately $\theta_{SH}$. Our study shows that the background noise distorts the measured signal even when the SNR is high enough, resulting in erroneous $\theta_{SH}$ values and may be one of the reasons for the large scattering in the values reported in the literature for common materials. Although the DLFMR is expected to be the most reliable method, there seems to be a fundamental limitation to the measurement. Our work marks another step towards utilizing the SHE in realistic practical applications.